\documentclass{svproc}
\usepackage[utf8]{inputenc}

\usepackage{url}

\usepackage{natbib}
\bibpunct{(}{)}{;}{a}{}{,}%
\usepackage{graphicx}
\usepackage{amsmath,amssymb}
\usepackage{multirow}
\usepackage{longtable}
\def\HI{H{\sc i}\, }

\def\Msun{$\textrm{M}_{\odot}$}

\def\HI{H{\sc i}\, }

\usepackage[b5paper]{geometry}
\begin{document}
	\mainmatter              
	\title{Observational insights on the formation scenarios of giant low surface brightness galaxies}
	\titlerunning{Observational insights on the formation of gLSBGs}  
	%
	\author{Anna Saburova\inst{1,2} \and Igor Chilingarian\inst{3,1} \and
		Anastasia Kasparova\inst{1}  \and Olga Sil'chenko\inst{1}  \and Ivan Katkov\inst{4,5,1}  \and Kirill Grishin\inst{1}  \and Roman Uklein\inst{6}  }
	\authorrunning{Anna Saburova et al.} 
	%
	\tocauthor{Anna Saburova,  Igor Chilingarian,
		Anastasia Kasparova,  Kirill Grishin, Olga Sil'chenko, Ivan Katkov, Roman Uklein}
	\institute{ Sternberg Astronomical Institute, M.V.~Lomonosov Moscow State University, Universitetsky pr., 13,  Moscow 119234, Russia,\\
		\email{saburovaann@gmail.com}
		\and
	Institute of Astronomy, Russian Academy of Sciences, Pyatnitskaya st., 48,  Moscow 119017, Russia
	\and
	Center for Astrophysics --- Harvard and Smithsonian, 60 Garden Street MS09, Cambridge, MA 02138, USA
	\and
	New York University Abu Dhabi, PO Box 129188 Abu Dhabi, UAE
	\and
	Center for Astro, Particle, and Planetary Physics, NYU Abu Dhabi, PO Box 129188, Abu Dhabi, UAE
	\and
	Special Astrophysical Observatory, Russian Academy of Sciences, Nizhniy Arkhyz, Karachai-Cherkessian Republic 357147, Russia}
	
	\maketitle

\begin{abstract}
	Giant low surface brightness galaxies (gLSBGs) with the disk radii of up to 130~kpc represent a challenge for currently accepted theories of galaxy formation and evolution, because it is difficult to build-up such large dynamically cold systems via mergers preserving extended disks. We summarize the in-depth study of the sample of 7 gLSBGs  based on the results of the performed spectral long-slit observations at the Russian 6-m BTA telescope of SAO RAS, surface photometry and \HI data available in literature. Our study revealed that most gLSBGs do not deviate from the Tully-Fisher relation. We discovered compact elliptical (cE) satellites in 2 out of these 7 galaxies. Provided the low statistical frequencies of gLSBGs and cEs, the chance alignment is improbable, so it can indicate that gLSBGs and cE are evolutionary connected and gives evidence in favor of the major merger formation scenario. Other formation paths of gLSBGs are also discussed. 
\end{abstract}

\paragraph{Introduction.}
 Giant low surface brightness galaxies (GLSBGs) deserve special interest since they harbor the largest rotating disks in the Universe. These galaxies have baryonic masses reaching $10^{11}$\Msun ~and dynamical masses $\sim10^{12}$\Msun. Currently accepted galaxy formation scenarios have hard time explaining their properties because for the mass assembly at such scale one typically needs dozens of minor and major mergers, which would likely destroy the disk, zero out the total angular momentum and turn a galaxy into a giant elliptical. Only a couple of dozens gLSBGs were found since the discovery of the prototypical galaxy Malin~1 in the 80s by \cite{Bothun1987} so they are considered extremely rare. The question of the gLSBGs formation remains open \citep{Kasparova2014, Galaz2015, Hagen2016, Boissier2016,Saburova2018, Saburova2021} despite their importance as a ``stress test'' for the $\Lambda$CDM cosmology. An in-plane major merger of two giant spiral galaxies with fine-tuned orbital parameters can end-up in a system resembling a gLSBG \citep{Saburova2018}. So does a triple major merger with two gas-rich galaxies when a host's hot halo gas cooling is triggered by the cold gas supply from the infalling systems and later forms a giant extended disc \citep{Zhuetal2018}. The analysis of the results of EAGLE simulations shows that the most extended LSB disks are formed by mergers \citep{Kulieretal2020}. Accretion of cold gas from a cosmic filament or gas-rich satellites \citep{Penarrubia2006} on-to a pre-existing high-surface brightness galaxy \citep{Saburova2019} can also foster a giant disk. The unusually sparse dark matter halo can also lead to the formation of giant LSB disc \citep{Kasparova2014}.
In this paper we briefly summarize the efforts that we made to understand the formation scenario of gLSBGs by observing a sample of 7 gLSBGs: Malin~1, Malin~2, NGC~7589, UGC~1378, UGC~1382, UGC~1922, and UGC~6614.
\paragraph{The results of long-slit spectral observations.}
We performed spectral long-slit observations at the 6-m BTA telescope of SAO RAS using SCORPIO and SCORPIO-2 spectrographs \citep{AfanasievMoiseev2005, AfanasievMoiseev2011}. The details of the data reduction and analysis could be found in \citet{Saburova2018,Saburova2019,Saburova2021}. The most prominent result of these observations is the discovered kinematically decoupled components in UGC~1922 and UGC~1382. In UGC~1382, the global gaseous disk counter-rotates with respect to stars, and this suggests the external origin of gas in the extended disk. Stellar populations in the central parts of 5 of the 7 gLSBGs are old and metal-rich supporting the scenario of accretion of gas onto a pre-existing high surface brightness galaxy.
\begin{figure}
\centering
\includegraphics[height=0.4\linewidth]{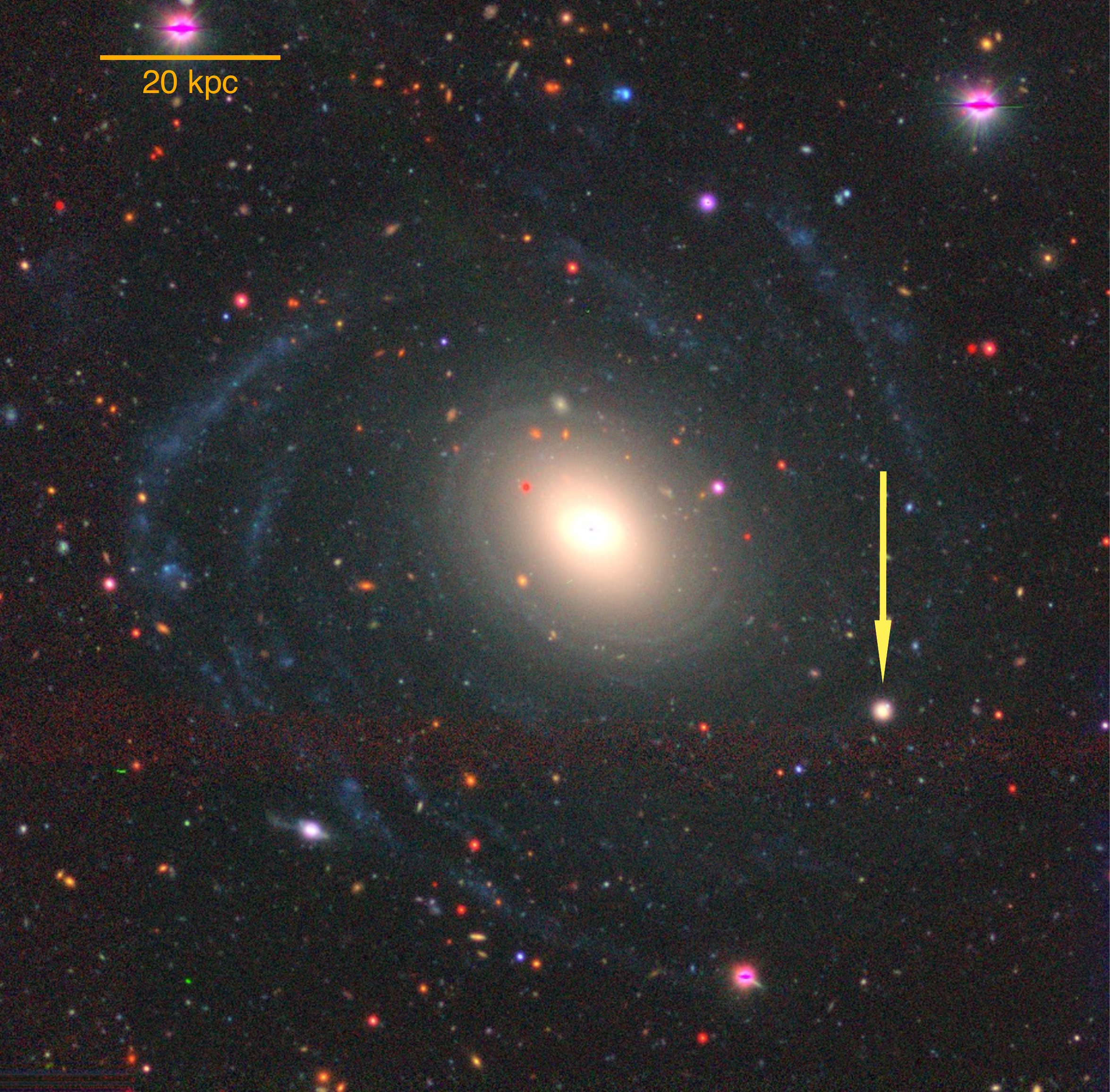}
\includegraphics[height=0.4\linewidth]{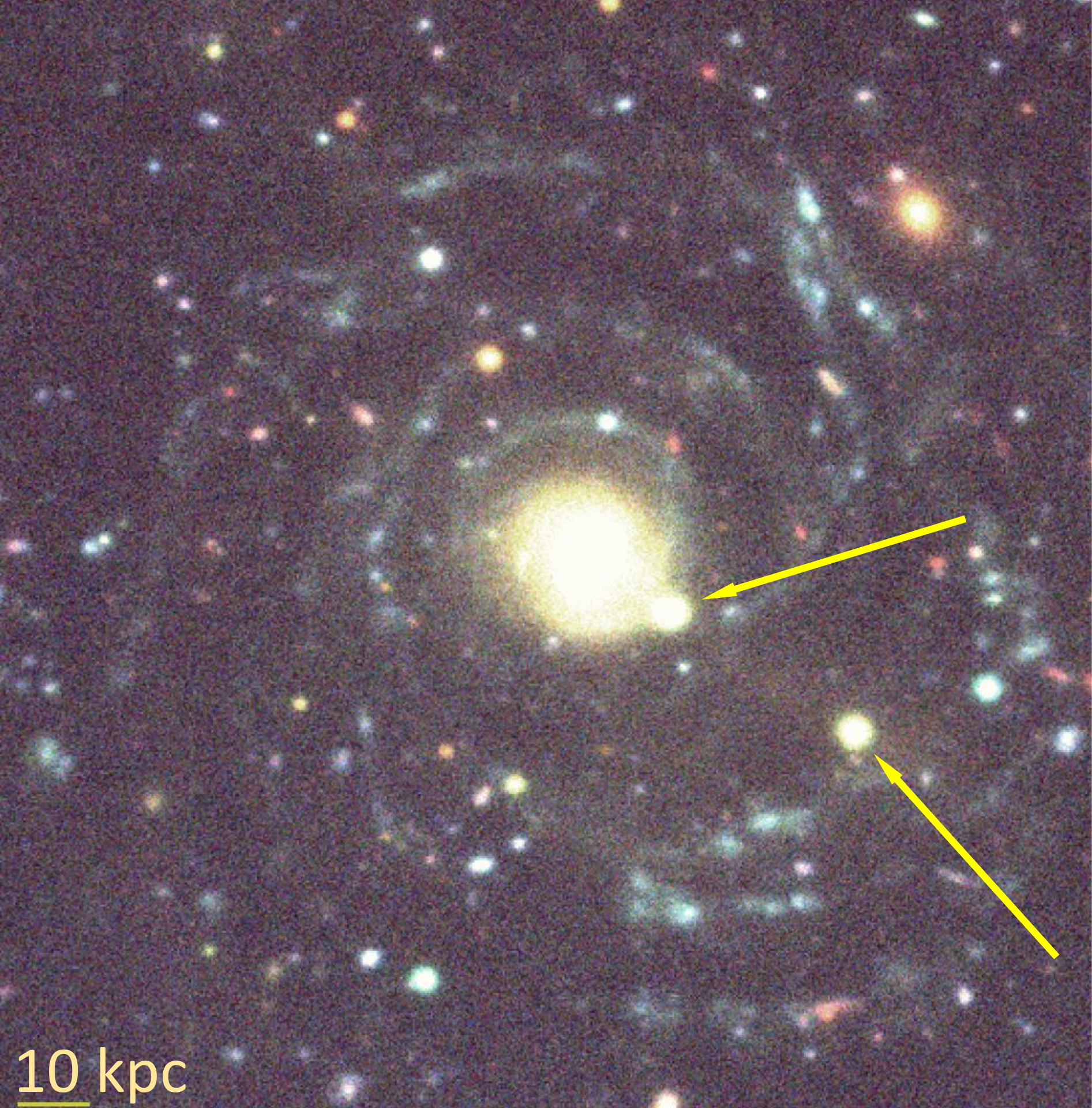}
\caption{The colour composite images of UGC~1382 from HSC (left panel) and Malin~1 (right panel) from the CFHT-Megacam Next Generation Virgo cluster Survey in $u$, $g$ and $i$-bands taken from \cite{Junais2019}. Arrows demonstrate the position of cE satellites}
\label{ce}
\end{figure}
\paragraph{The discovery of the cE satellites.}
The inspection of the archival HST images of the class prototype Malin~1 allowed us to make an important step towards understanding of the nature of gLSBGs. Malin~1 appeared to have two satellites which reminded the low-mass dense compact elliptical galaxies (cE) (see Fig. \ref{ce}). The cEs are rare low-mass systems ($\sim0.5$\% of dwarf galaxy population) with small sizes ($r_e<1$~kpc) and high densities typically hosting old metal-rich stars \cite{Chilingarianetal2009, 2021arXiv210309241F}. They likely originate from tidal stripping \cite{2001ApJ...552L.105B} of massive disky progenitors losing 90--95\% of stellar mass \cite{Chilingarianetal2009, 2009MNRAS.397.1816P, 2014MNRAS.443.1151N, ChilingarianZolotukhin2015,2021arXiv210309241F}. The structural analysis of the Hubble WFPC2 data using GALFIT confirmed the cE classification, and the re-analysis of published optical spectra from the Russian 6-m telescope confirmed that these cE were indeed the satellites of Malin~1. Further visual inspection of high-quality optical image of UGC~1382 from Subaru Hyper-SuprimeCam Strategic survey revealed  additional cE-satellite of gLSBG, the SDSS spectral data confirmed the physical association. As the next step we significantly extended our sample of gLSBGs by visual inspection of HSC and DECaLS data (Saburova et al. in prep.) and discovered more cE+gLSBGs systems (Chilingarian et al. in prep.).  Given the statistical frequencies of gLSBGs and cEs, the chance of alignment is improbable thus  cE and gLSBGs must be evolutionary connected which gives more arguments in favor of merger scenario at least for some of gLSBGs.
\paragraph{The baryonic Tully-Fisher relation}
 \citep[BTFR][]{Sprayberry1995,McGaughSchombert2015}  between gas+stellar mass and rotation velocity of galaxies is a useful tool for the diagnostic of the formation paths of galaxies. We found out that 6 out of the 7 gLSBGs lie on the high-mass extension of the BTFR which is also the case for the most luminous spiral galaxies \citep{DiTeodoro2021}. This suggests that gLSBGs are scaled-up versions of less massive disks.
 \paragraph{The parameters of dark matter halos} of gLSBGs that we were able to derive from \HI+ optical rotation curves show dichotomy. Some gLSBGs have high radial scales of the dark halo density profile, while some others do not. Hence, the gLSBG class is in fact inhomogeneous, and there might exist different formation channels. At the same time according to our findings \citep{Saburova2021}, most gLSBGs have external origin of the material for their extended disks (merger and accretion scenarios). 
 
 \textbf{Acknowledgements:}  The mass modelling of the rotation curves was done with the support of the Russian Science Foundation (RScF) grant No. 19-72-20089.
\bibliographystyle{aa}
\bibliography{template}
\end{document}